\documentclass[aps,pre,twocolumn,showpacs,longbibliography]{revtex4-1}
\usepackage{bbding}
\usepackage{amsmath}
\usepackage{amsthm}
\usepackage{graphicx}
\usepackage{subfigure}
\usepackage{amssymb}
\usepackage{graphicx}
\usepackage{dcolumn}
\usepackage{bm}
\usepackage{float}
\usepackage[colorlinks,linkcolor=blue,citecolor=blue,urlcolor=blue,hyperindex,breaklinks]{hyperref}

\begin{document}

\title{Entanglement enhanced and one-way steering in $\mathcal{PT}$%
-symmetric cavity magnomechanics}
\author{Ming-Song Ding}
\author{Chong Li}
\email{lichong@dlut.edu.cn}
\affiliation{School of Physics, Dalian University of Technology, Dalian
116024, China}
\date{\today }

\begin{abstract}
We study creation of entanglement and quantum steering in a parity-time- ($%
\mathcal{PT}$ -) symmetric cavity magnomechanical system. There is magnetic
dipole interaction between the cavity and photon-magnon, and there is also
magnetostrictive interaction which is induced by the phonon-magnon coupling
in this system. By introducing blue-detuned driving microwave field to the
system, the bipartite entanglement of the system with $\mathcal{PT}$
-symmetry is significantly enhanced versus the case in the conventional
cavity magnomechanical systems (loss-loss systems). Moreover, the one-way
quantum steering between magnon-phonon and photon-phonon modes can be
obtained in the unbroken-PT -symmetric regime. The boundary of stability is
demonstrated and this show that the steady-state solutions are more stable
in the gain and loss systems. This work opens up a route to explore the
characteristics of quantum entanglement and steering in magnomechanical
systems, which might have potential applications in quantum state
engineering and quantum information.
\end{abstract}

\maketitle

\section{Introduction}

In recent years, cavity magnomechanical system (CMM system) has attracted
extensive attentions in the cavity quantum electrodynamics. This CMM system
consists of a three dimensional rectangular microwave cavity and a
single-crystal yttrium-iron-garnet (YIG) sphere inside. Owing the high spin
density and the strong spin-spin exchange interactions, the Kittel mode in
the YIG sphere can achieve strong \cite{k1,k2} and even ultrastrong coupling 
\cite{k201} to the microwave cavity mode. And this strong coupling can be
achieved even at room temperature \cite{k3}. On the other hand, the CMM
systems are developed from optomechanical systems \cite%
{k4,k5,k501,k502,k49,k503,k2} which achieve the interaction between phonons
and optical or microwave photons by radiation force or electrostatic force.
While the magnetostrictive force of the YIG sphere is applied to realize the
coupling between phonons and magnons in the CMM systems. Comparing with the
optomechanical systems, the CMM systems have the advantages of high
adjustability and low loss. Therefore, it provides a good opportunity for
realizing highly tunable information processing in the hybrid quantum
systems \cite{k61}. J. Q. You $et$ $al$. report that the bistability of
cavity magnon polaritons \cite{k62}, G. S. Agarwal $et$ $al$. show that the
tripartite entanglement among magnons, photons, and phonons \cite{k7}.
Besides, the high-order sideband generation\cite{k9}, magnon Kerr effect\cite%
{k8}, the light transmission in cavity-magnon system \cite{k20} and others
are also studied \cite{k21,k11,k22,k221,k222,k223,k224,k225,k226}.

The developments in Parity-time-symmetry ($\mathcal{PT}$-symmetry) optical
structure resulted in the birth of the new field which attracted
considerable interest \cite{k12,k13,k14}. In the past, people believed that
only Hermitian Hermitonian has real eigenvalue spectra, while C. M. Bender $%
et$ $al$. have proved that the $\mathcal{PT}$-symmetric non-Hermitian
Hamiltonian ($[H,PT]=0$) can also has real eigenvalue spectra \cite{k13}. Up
to present, $\mathcal{PT}$-symmetry been widely applied to quantum optics,
quantum information processing, and propagation, including the optical
non-reciprocity in $\mathcal{PT}$-symmetric whispering-gallery microcavities
(WGM) \cite{k18}, the detection sensitivity of weak mechanical motion \cite%
{k19}, realizing quantum chaos \cite{k17}, strengthening optics nonlinearity%
\cite{k16}, nonreciprocal light propagation \cite{k161} and so on \cite%
{k15,k23,k231,k32,k321}. In addition, it is difficult to achieve ideal $%
\mathcal{PT}$-symmetry under the strict requirements of balanced gain and
loss. Here, we study the non-equilibrium effective $\mathcal{PT}$-symmetric
system, which works in microwave regime.

In this work, we propose to construct a $\mathcal{PT}$-symmetric CMM system
with the active cavity and passive magnon modes. The magnetostrictive
(radiation pressure like) interaction mediates the coupling between magnons
and phonons. And the cavity photons and magnons are coupled via magnetic
dipole interaction. We show the stability parameter boundary of the system
which is driven by a blue-detuned microwave field. Here, $\mathcal{PT}$%
-symmetry leads to a strong bipartite entanglement among the mechanical
mode, the optical field inside the gain cavity and magnon mode, the
logarithmic negativity is used to measure the continuous variable (CV)
entanglement \cite{k43}. And the potential feasibility of the experiment is
discussed. G. S. Agarwal $et$ $al$. first studied the entanglement in CMM
system \cite{k7}, our work is based on it and consider the enhancement
effect of $\mathcal{PT}$-symmetry on the entanglement. Furthermore, we show
that the $\mathcal{PT}$-symmetry can induce one-way quantum steering between
magnon-phonon and photon-phonon modes. As we know, the quantum steering is
intrinsically different from quantum entanglement and Bell nonlocality for
its asymmetric characteristics and it has potential applications in the
quantum information protocols, such as device-independent quantum key
distribution.

The structure of the paper is as follows. In Sec. II, we introduce the
Hamiltonian and dynamical equations of the whole systems. In Sec. III, the
stability of the system is discussed. In Sec. IV, we show that by
introducing $\mathcal{PT}$-symmetry, the entanglement is obviously enhanced
and one-way steering can be obtained. Finally, a concluding summary is given.

\section{Model and \protect\bigskip dynamical Hamiltonian}

We utilize a $\mathcal{PT}$-symmetric CMM system as shown in Fig. 1(a),
which consists of microwave cavity photons (gain) and magnons (loss). The
magnons in the YIG sphere are collective excitation of magnetization, and
the uniform magnon mode is driven by an adjustable microwave field.
Furthermore, the magnetic dipole interaction leads to the coupling between
magnon mode and active cavity mode.

Owing to the magnetostrictive effect, the YIG sphere can be considered as an
excellent mechanical resonator \cite{k61}. Therefore, the term of coupling
between magnons and phonons can be introduced into Hamiltonian of the
system. And the magnetostrictive coupling strength is determined by the mode
overlap between the magnon and phonon modes. In general, the magnomechanical
coupling is very weak \cite{k61}. However, it can be effectively enhanced by
a microwave driving field \cite{k62,k7}.

The equivalent mode-coupling model is given in Fig. 1(b). The size of the
YIG sphere we considered is much smaller than the wavelength of the
microwave field. Hence, the interaction between microwave cavity photons and
phonons induced by the radiation pressure is neglected. In a rotating-wave
approximation, the Hamiltonian of the whole system is given by ($\hbar =1$) 
\cite{k7}

\begin{eqnarray}
H &=&\omega _{a}a^{\dagger }a+\omega _{m}m^{\dagger }m+g_{ma}(a^{\dagger
}m+m^{\dagger }a)  \notag \\
&&+\frac{\omega _{b}}{2}(x^{2}+p^{2})+g_{mb}m^{\dagger }mx  \label{eq01} \\
&&+i\varepsilon _{d}(m^{\dagger }e^{-i\omega _{d}t}-me^{i\omega _{d}t}), 
\notag
\end{eqnarray}%
where $a(a^{\dagger })$ and $m(m^{\dagger })$ are the annihilation(creation)
operators of the cavity mode and the uniform magnon mode at the frequency $%
\omega _{a}$ and $\omega _{m}$, respectively ($[O,O^{\dagger }]=1,O=a,m$).
The magnon frequency $\omega _{m}$ can be easily adjusted by altering the
external bias magnetic field $H$ via $\omega _{m}=\gamma _{g}H$, where $%
\gamma _{g}$ is the gyromagnetic ratio ($\gamma /2\pi =28GHz/T$). $x$ and $p$
are the dimensionless position and momentum quadrature of the mechanical
mode with the frequency $\omega _{b}$ ($[x,p]=1$). The coupling rate of the
magnon-cavity interaction is $g_{ma}$ and $\omega _{d}$ is frequency of
driven field. Here, the Hamiltonian in Eq.(\ref{eq01}) does not include the
gain.

\begin{figure}[tbp]
\centering\includegraphics[width=7cm]{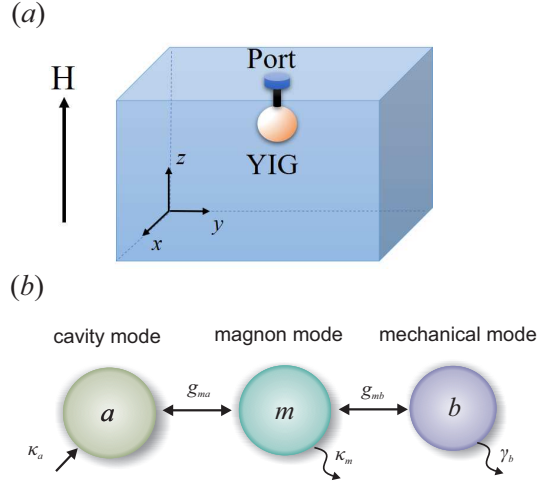} % If not, use
%\vspace{5cm}       % Give the correct figure height in cm
% Give a unique label
\caption{(a) Schematic illustration of the system, a YIG sphere is placed in
a three-dimensional cavity, and it is connected with the inner wall of the
cavity by a silicon fiber. The port is for driving the YIG sphere via a
microwave field. A bias magnetic field $H$ along the $z$-axis is used to
realize magnon-photon coupling by magnetic dipole interaction. It is worth
mentioning that the directions of the bias magnetic field, the drive field,
and the magnetic field of the cavity mode are perpendicular to each other.
(b) The equivalent mode-coupling model.}
\end{figure}

Under the assumption of the low-lying excitations, the Rabi frequency $%
\varepsilon _{d}$ is $\frac{\sqrt{5}}{4}\gamma _{g}\sqrt{N_{t}}B_{0}$, and
it denotes the coupling strength between the magnon mode and the driving
field with the amplitude $B_{0}$. The total number of the spins $N_{t}$ is
given by $\rho V$, where the spin density $\rho =$ $4.22\times 10^{27}$m$%
^{-3}$, and $V$ is the volume of the YIG sphere.

After a rotating frame at the frequency $\omega _{d}$ of the above
Hamiltonian, one derives the following set of the Heisenberg-Langevin
equations:

\begin{eqnarray}
\dot{a} &=&(-i\Delta _{a}+\kappa _{a})a-ig_{ma}m+\sqrt{2\kappa _{a}}a^{in}, 
\notag \\
\dot{m} &=&(-i\Delta _{m}-\kappa _{m})m-ig_{ma}a-ig_{mb}mx+\varepsilon _{d}+%
\sqrt{2\kappa _{m}}m^{in},  \notag \\
\dot{x} &=&\omega _{b}p,  \label{eq02} \\
\dot{p} &=&-\omega _{b}x-\gamma _{b}p-g_{mb}m^{\dagger }m+\xi ,  \notag
\end{eqnarray}%
where the detuning $\Delta _{a(m)}=\omega _{a(m)}-\omega _{d}$, $\kappa _{a}$
is the gain rate of the cavity mode, $\kappa _{m}$ and $\gamma _{b}$ are the
dissipation rates of magnon and mechanical modes, respectively. $%
a^{in},m^{in}$ and $\xi $ are input noise operators of cavity, magnon and
mechanical modes. With a Markovian approximation, the input noise
correlation functions are shown as: $\left\langle a^{in}(t)a^{in\dagger
}(t^{\prime })\right\rangle =(n_{a}+1)\delta (t-t^{\prime })$, $\left\langle
a^{in\dagger }(t)a^{in}(t^{\prime })\right\rangle =n_{a}\delta (t-t^{\prime
})$, $\left\langle m^{in}(t)m^{in\dagger }(t^{\prime })\right\rangle
=(n_{m}+1)\delta (t-t^{\prime })$, $\left\langle m^{in\dagger
}(t)m^{in}(t^{\prime })\right\rangle =n_{m}\delta (t-t^{\prime })$, $%
\left\langle \xi (t)\xi ^{\dagger }(t^{\prime })\right\rangle
=(n_{b}+1)\delta (t-t^{\prime })$ and $\left\langle \xi ^{\dagger }(t)\xi
(t^{\prime })\right\rangle =n_{b}\delta (t-t^{\prime })$. Here, $n_{\mu
}=(e^{\hbar \omega _{\mu }/k_{B}T}-1)$($\mu =a,m,b$) with $k_{B}$ the
Boltzmann constant and $T$ the environmental temperature, and these $n_{\mu
} $ are equilibrium mean thermal photon, magnon, and phonon numbers,
respectively.

In order to better understand the broken $\mathcal{PT}$-symmetry regimes and
the unbroken $\mathcal{PT}$-symmetry regimes of this system, we only focus
on the cavity and magnon modes. And the driving field in Eq.(\ref{eq01}) can
be neglected as the same reason in \cite{k18}. By the $\mathcal{PT}$
operation, the Hamiltonian can be described by a second-order matrix, i.e.,

\begin{equation}
H_{\mathcal{PT}}=\mathcal{PT}H\mathcal{PT}=(\hat{a}^{\dagger }\text{ }\hat{m}%
^{\dagger })\left( 
\begin{array}{cc}
\Delta _{m}+i\kappa _{m} & g_{ma} \\ 
g_{ma} & \Delta _{a}-i\kappa _{a}%
\end{array}%
\right) (\hat{a}\text{ }\hat{m})^{T},  \label{eq03}
\end{equation}%
where the parity operation $\mathcal{P}$ acting on the Hamiltonian can
interchange the loss and gain of the cavity and magnon modes, i.e., $\hat{a}%
\Longleftrightarrow -\hat{m}$ and $\hat{a}^{\dagger }\Longleftrightarrow -%
\hat{m}^{\dagger }.$And the time reversal operation $\mathcal{T}$ on $H$ can
reverse the sign of complex number $i$. After setting $\Delta _{a}=\Delta
_{m}=\Delta $, the eigenfrequencies of the Hamiltonian in Eq.(\ref{eq03})
can be written as

\begin{equation}
\omega _{\pm }=-\Delta -i(\kappa _{m}-\kappa _{a})/2\pm \sqrt{%
g_{ma}^{2}-(\kappa _{a}+\kappa _{m})^{2}/4}.  \label{eq04}
\end{equation}

In order to make the Hamiltonian be $\mathcal{PT}$-symmetry, the
eigenfrequencies should be real. According to Eq.(\ref{eq04}), with the the
condition: $\Delta _{a}=\Delta _{m}=\Delta $, $2g_{ma}>\kappa _{a}+\kappa
_{m}$ and $\kappa _{m}=\kappa _{a}$, we have $H^{\mathcal{PT}}=H$ and $[H,%
\mathcal{PT]}=0$, that is to say, this Hamiltonian is strictly in unbroken $%
\mathcal{PT}$-symmetry regime. Correspondingly, the broken-$\mathcal{PT}$%
-symmetry regime holds for the case of $2g_{ma}<\kappa _{a}+\kappa _{m}$. In
addition, the phase transition between the broken-$\mathcal{PT}$-symmetry
and unbroken $\mathcal{PT}$-symmetry regimes, i.e., $2g_{ma}=\kappa
_{a}+\kappa _{m}$ is exceptional point (EP).

It is worth noting that under the condition of non-equilibrium ($\kappa
_{m}\neq \kappa _{a}$), even if the eigenvalues of the system are complex,
the system still has phase transition, and phase transformation point
remains unchanged. Actually, choosing appropriate reference frame, the
non-equilibrium system is an effective $\mathcal{PT}$-symmetric system.
Physically, this system can be considered as a strict $\mathcal{PT}$%
-symmetric system coupled to an effective reservoir with decay rate $\kappa
_{m}-\kappa _{a}$. A lot of work have adopted the effective $\mathcal{PT}$%
-symmetric systems\cite{k12,k121,k222}.

Next, the values of the specific parameters used in this work are given and
they are easy to be achieved in experiments \cite{k61,k62}. In our
discussion, $\omega _{a}/2\pi =\omega _{m}/2\pi =10.1GHz$, $\omega _{b}/2\pi
=10MHz$, $\gamma _{b}/2\pi =10Hz$, $g_{mb}/2\pi =0.2Hz,\kappa _{m}=1MHz$,
and the temperature is $20$mK.

\section{STABILITY OF SYSTEM}

In this section, in order to quantify the entanglement of this system, an
important condition is the existence of asymptotic steady state and system
will keep the state for a long evolution time. Hence, we discuss the
stability of the steady-state solutions.

Because the magnon mode is directly driven by a strong microwave source, it
leads to a large number of magnons $\left\vert \left\langle m\right\rangle
\right\vert \gg 1$ at the steady state. And according to the cavity-magnon
beam splitter interaction in Eq.(\ref{eq04}), the cavity field has a large
amplitude $\left\vert \left\langle a\right\rangle \right\vert \gg 1$.
Therefore, each Heisenberg operator can rewritten as a sum of its
steady-state mean value and its corresponding quantum fluctuation, i.e., $%
O(t)=O_{s}+\delta O(t)$ $(O=a,m,x,p)$. Then we study the stability of the
system through a linear stability analysis \cite{k37}. From the
Heisenberg-Langevin equations in Eq.(\ref{eq02}), the dynamical equation of
the quantum fluctuation is written by a compact equation, i.e.,

\begin{equation}
\dot{\nu}(t)=M\nu (t)+r(t),  \label{eq05}
\end{equation}%
where $v(t)$ is the vector of the quantum fluctuations, and $r(t)$ is the
noise vector. They can be expressed as $v(t)=[\delta a(t)$, $\delta
a^{\dagger }(t)$, $\delta m(t)$, $\delta m^{\dagger }(t)$, $\delta x(t)$, $%
\delta p(t)]^{T}$ and $r(t)=[\sqrt{2\kappa _{a}}\delta a^{in}(t)$, $\sqrt{%
2\kappa _{a}}\delta a^{in\dagger }(t)$, $\sqrt{2\kappa _{m}}\delta m^{in}(t)$%
, $\sqrt{2\kappa _{m}}\delta m^{in\dagger }(t)$, $0$, $\xi (t)]^{T}$,
respectively. The matrix $M$ is the coefficient matrix of the system, which
reflects the stability or stochastic property of the system. Here, $M$ can
be obtained as

\begin{equation}
\left( 
\begin{array}{cccccc}
-i\Delta _{a}+\kappa _{a} & 0 & -ig_{ma} & 0 & 0 & 0 \\ 
0 & i\Delta _{a}+\kappa _{a} & 0 & ig_{ma} & 0 & 0 \\ 
-ig_{ma} & 0 & -i\tilde{\Delta}_{m}-\kappa _{m} & 0 & -iG & 0 \\ 
0 & ig_{ma} & 0 & i\tilde{\Delta}_{m}-\kappa _{m} & iG & 0 \\ 
0 & 0 & 0 & 0 & 0 & \omega _{b} \\ 
0 & 0 & -G & -G & -\omega _{b} & -\gamma _{b}%
\end{array}%
\right) ,  \label{eq08}
\end{equation}%
where $G=g_{mb}\left\vert m_{s}\right\vert $ is the
coherent-driving-enhanced magnomechanical coupling strength, $m_{s}$ can be
obtained by solving the steady-state mean value of Eq.(\ref{eq02}), i.e.,

$\ \ \ \ \ \ \ \ \ \ \ \ \ \ \ \ \ \ \ \ \ $%
\begin{equation}
m_{s}=\frac{\varepsilon _{d}(i\Delta _{a}-\kappa _{a})}{g_{ma}^{2}+(i\Delta
_{a}-\kappa _{a})(i\tilde{\Delta}_{m}+\kappa _{m})},  \label{eq09}
\end{equation}%
$\bigskip $where $\tilde{\Delta}_{m}=\Delta _{m}+g_{mb}\left\vert
q_{s}\right\vert $ is the effective magnon-drive detuning. 

The stability analysis of the system can be done according to the
eigenvalues of the matrix $M$, it can be seen that the matrix $M$ has three
pairs of conjugate eigenvalues. And the real parts of the eigenvalues are
known as the Lyapunov exponents \cite{k38}, if the maximal Lyapunov exponent
is negative, the system is stable. Contrarily, the maximal Lyapunov exponent
is positive indicates the system is unstable \cite{k39,k40}.

In Figs.2 (a), (b) and (c), the boundary of linear stability is shown. The
light area indicates that the maximal Lyapunov exponent is negative, in
other words, the system is stable. The remaining dark area indicates that
the system is unstable. Here, we use $\kappa _{a}>0$ and $\kappa _{a}<0$ to
represent the active-passive CMM system and the conventional CMM system,
respectively. Comparison of Figs.2(a) and 2(b), the stable area of (b) $%
\kappa _{a}=0.2\kappa _{m}$ is obviously larger than that of (a) $\kappa
_{a}=-0.2\kappa _{m}$. It means that the stability of active-passive CMM
system ($\kappa _{a}>0$) is better than that of the conventional CMM system (%
$\kappa _{a}<0$) within the range of parameters we considered. Then from
Fig.2(c), it can be conclude that the stability of the active-passive CMM
system decreases with the system approaches the gain-loss balance. Finally,
it can be seen that the stability can be improved by the increase of $g_{ma}$
in the active-passive CMM system in Fig.2(d).

When the active-passive CMM system is in the EP or broken $\mathcal{PT}$%
-symmetry regime, in order to get a stable system, $G$ needs to be small
enough compared to $g_{ma}$, which leads to a very weak entanglement (it
will be discussed in detail in the next section). As a consequence, we only
focus on the stability of unbroken $\mathcal{PT}$-symmetry regime in this
section.

\begin{figure}[tbp]
\centering% Use the relevant command for your figure-insertion program
\resizebox{0.5\textwidth}{!}{  \includegraphics{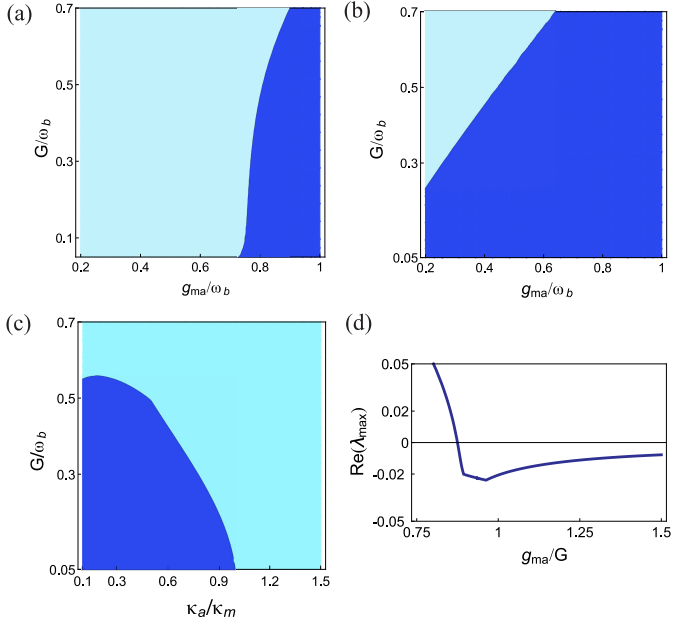}
} % If not, use
%\vspace{5cm}       % Give the correct figure height in cm
% Give a unique label
\caption{(a), (b) and (c) The boundary of linear stability. The light(dark)
areas stands for the evolution of the system is unstable(stable). (a) The
conventional CMM system $\protect\kappa _{a}=$ $-0.2\protect\kappa _{m}$;
(b) the active-passive CMM system $\protect\kappa _{a}=$ $0.2\protect\kappa %
_{m}$. (c) The stability boundary as the system closer to the balanced gain
and loss limit. $g_{ma}=0.5\protect\omega _{b}$. (d) The maximal Lyapunov
exponent as function of the ratio of $g_{ma}/G$ ($G$ is fixed) $G=0.4\protect%
\omega _{b}$, $\protect\kappa _{a}=$ $0.2\protect\kappa _{m}$. The
parameters we used are $\protect\omega _{b}/2\protect\pi =10MHz$, $\Delta
_{a}=\tilde{\Delta}_{m}=-\protect\omega _{b}$, $\protect\kappa _{m}=0.1%
\protect\omega _{b}$.}
\end{figure}

\section{Entanglement And Steering}

Here, we choose the logarithmic negativity $E_{\mathcal{N}}$ to measure the
bipartite entanglements of the system \cite{k43}. $E_{\mathcal{N}}$ can be
computed by the quantum fluctuations of the system's quadratures \cite{k41}.
Then the quadratures of the quantum fluctuations are introduced, the vector
of quadratures is given by $u(t)=[\delta X_{1},\delta X_{2},\delta
Y_{1},\delta Y_{2},\delta x,\delta p]^{T}$ and the vector of noise is $n(t)=[%
\sqrt{2\kappa _{a}}X_{1}^{in}(t),\sqrt{2\kappa _{a}}X_{2}^{in}(t),\sqrt{%
2\kappa _{m}}y_{1}^{in}(t),\sqrt{2\kappa _{m}}y_{2}^{in}(t),0,\xi (t)]^{T}$,
where $\delta X_{1}=(\delta a+\delta a^{\dagger })/\sqrt{2}$, $\delta
X_{2}=i(\delta a^{\dagger }-\delta a)/\sqrt{2}$, $\delta Y_{1}=(\delta
m+\delta m^{\dagger })/\sqrt{2}$and $\delta Y_{2}=i(\delta m^{\dagger
}-\delta m)/\sqrt{2}$. Similarly, the input noise quadratures are defined in
the same way.

Eq.(\ref{eq02}) can be written in a compact form $\dot{u}(t)=A\nu (t)+n(t)$,
where the correlation matrix

\begin{equation}
A=\left( 
\begin{array}{cccccc}
\kappa _{a} & \Delta _{a} & 0 & g_{ma} & 0 & 0 \\ 
-\Delta _{a} & \kappa _{a} & -g_{ma} & 0 & 0 & 0 \\ 
0 & g_{ma} & -\kappa _{m} & \tilde{\Delta}_{m} & -G & 0 \\ 
-g_{ma} & 0 & -\tilde{\Delta}_{m} & -\kappa _{m} & 0 & 0 \\ 
0 & 0 & 0 & 0 & 0 & \omega _{b} \\ 
0 & 0 & 0 & G & -\omega _{b} & -\gamma _{b}%
\end{array}%
\right) .  \label{eq10}
\end{equation}

Due to the dynamics of the system is linearized and the input noise
operators $\xi $, $m^{in}$ and $a^{in}$ are Gaussian noise, the steady-state
of the quantum fluctuations is a continuous variable (CV) three-mode
Gaussian state. It can be obtained by a $6\times 6$ steady-state covariance
matrix (CM) $V$, which can be solved by the Lyapunov equation, i.e., \cite%
{k41}

\begin{equation}
AV+VA^{T}=-D,  \label{eq11}
\end{equation}%
where the elements of the diffusion matrix $D$ are defined by $\delta
(t-t^{^{\prime }})D_{i,j}=\left\langle n_{i}(t)n_{j}(t^{^{\prime
}})+n_{j}(t^{^{\prime }})n_{i}(t)\right\rangle /2$. From the input noise
correlation functions, one obtains $D=diag[\kappa _{a}(2n_{a}+1)$, $\kappa
_{a}(2n_{a}+1)$, $\kappa _{m}(2n_{m}+1)$, $\kappa _{m}(2n_{m}+1)$, $0$, $%
\gamma _{b}(2n_{b}+1)]$.

In the CV case, $E_{\mathcal{N}}$ can be defined as

\begin{equation}
E_{\mathcal{N}}=\max [0,-\ln 2\eta ^{-}],  \label{eq13}
\end{equation}%
where $\eta ^{-}=2^{-1/2}\{\Sigma (V)-[\Sigma (V)^{2}-4\det
V_{s}]^{1/2}\}^{1/2}$, with $\Sigma (V)=\det A+\det B-2\det C$. Here, $V_{s}$
is a reduced $4\times 4$ submatrix for the covariance matrix (CM). And the
matrix elements of $V_{s}$ depend on the pairwise entanglement of two
interesting modes (the photon-magnon, magnon-phonon and photon-phonon
modes), it can be rewritten as 
\begin{equation}
V=\left( 
\begin{array}{cc}
A & C \\ 
C^{T} & B%
\end{array}%
\right) .  \label{eq14}
\end{equation}

As we know, the quantum steering is different from the entanglement for it
has asymmetric characteristics between the parties. For the Gaussian states
of the two interesting modes, a Gaussian quantum steering based on the form
of quantum coherent information is introduced \cite{k48}. Here, the $\chi
_{1}\rightarrow $ $\chi _{2}$ steering is given by

\begin{equation}
S_{\chi _{1}\rightarrow \chi _{2}}=\max [0,\frac{1}{2}\ln \frac{\det A}{%
4\det V_{s}}],  \label{eq15}
\end{equation}%
where $\chi _{1}$ and $\chi _{2}$ stand for two interesting modes. A
corresponding measure of Gaussian $\chi _{2}$ $\rightarrow $ $\chi _{1}$
steerability can be obtained by swapping $\chi _{1}$ and $\chi _{2}$, which
can be expressed as

\begin{equation}
S_{\chi _{2}\rightarrow \chi _{1}}=\max [0,\frac{1}{2}\ln \frac{\det B}{%
4\det V_{s}}].  \label{eq16}
\end{equation}

\begin{figure}[tbp]
\centering% Use the relevant command for your figure-insertion program
\resizebox{0.55\textwidth}{!}{  \includegraphics{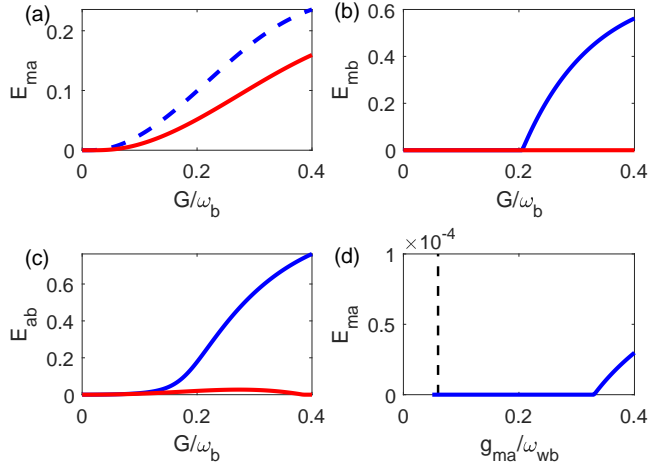}
} % If not, use
%\vspace{5cm}       % Give the correct figure height in cm
% Give a unique label
\caption{(a), (b) and (c) The entanglements $E_{\mathcal{N}\text{,}am}$, $E_{%
\mathcal{N}\text{,}bm}$ and $E_{\mathcal{N}\text{,}ab}$ as functions of the
effective coupling rates $G$. The blue dotted line denotes the case of $%
\protect\kappa _{a}=0.2\protect\kappa _{m}$ (the $\mathcal{PT}$-symmetry CMM
system) and the red solid line denotes the case of $\protect\kappa _{a}=-0.2%
\protect\kappa _{m}$ (the conventional CMM system). (d) The entanglements $%
E_{\mathcal{N}\text{,}am}$ as function of the ratio of $\protect\kappa _{a}/%
\protect\kappa _{m}(\protect\kappa _{m}$ is fixed$)$. We set $g_{ma}=\protect%
\omega _{b}$, the other parameters we chose are the same as those in Fig.2}
\end{figure}

From Fig. 3(a), (b) and (c), the bipartite entanglements $E_{\mathcal{N}%
\text{,}am}$, $E_{\mathcal{N}\text{,}bm}$ and $E_{\mathcal{N}\text{,}ab}$
are significantly enhanced by $\mathcal{PT}$-symmetry compared to what is
generated in a conventional CMM system, where $E_{\mathcal{N}\text{,}am}$, $%
E_{\mathcal{N}\text{,}bm}$ and $E_{\mathcal{N}\text{,}ab}$ denote the
cavity-magnon, magnon-phonon, and cavity-phonon entanglement. From Fig.
3(a), we find that with the introduction of the magnomechanical interaction,
the directly coupled photons and magnons begin to entangle. In Fig. 3(b), it
can be seen that the entanglement $E_{\mathcal{N}\text{,}bm}$ in the
conventional CMM system is $0$. The reason is that the magnon mode driven by
the blue-detuning can not cause the anti-Stokes process, which cools the
mechanical mode. That is to say, the phonons cannot be cooled by the
magnomechanical interaction in the conventional CMM system, thus it hinders
the generation of entanglement. However, for $\mathcal{PT}$-symmetric CMM
system we considered, the red-detuning driving field lead to the instability
of the system. Hence, the case of blue-detuning is chosen and the strong
entanglement can still be obtained in unbroken $\mathcal{PT}$-symmetry
regime. Fig. 3(d) shows $E_{\mathcal{N}\text{,}am}$ versus $g_{ma}$ in the
active-passive CMM system. The black vertical dotted line represents EP,
which corresponds to $g_{ma}=(\kappa _{a}+\kappa _{m})/2=0.06\omega _{b}$.
The entanglement occurs in the unbroken $\mathcal{PT}$-symmetry regime, and
there is no entanglement in EP and broken $\mathcal{PT}$-symmetry regime. In
order to make the system stable, we set $G=0.03\omega _{b}$, which leads to
a very weak entanglement. In addition, the areas without solid blue lines
represent that the system has no steady state.

In this work, all the results satisfy the stability conditions mentioned in
the previous section. The effective magnomechanical coupling $G=4MHz$
corresponding to the drive power $27.7$mW, at the drive magnetic field $%
B_{0}\approx 6.88\times 10^{-5\text{ }}$and $g_{mb}=0.4Hz$ \cite{k7}. For
this system, $G$ can be changed by adjusting the the drive power. In
addition, we set $G>\kappa _{m}$ to ensure that unwanted magnon Kerr effect
can be neglected in a strong magnon driving field \cite{k8,k62}.

\begin{figure}[tbp]
\centering% Use the relevant command for your figure-insertion program
\resizebox{0.55\textwidth}{!}{  \includegraphics{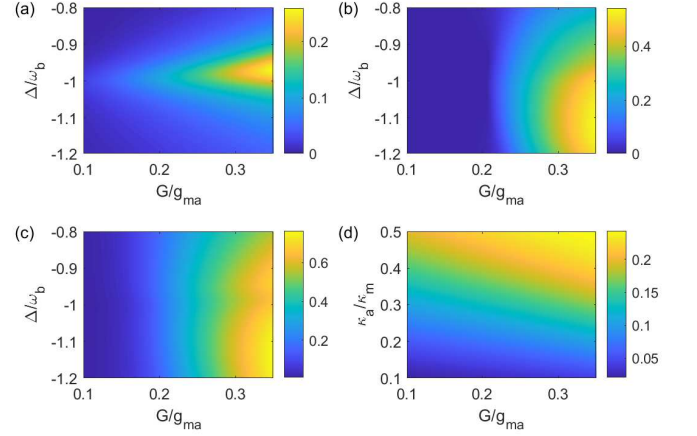}
} % If not, use
%\vspace{5cm}       % Give the correct figure height in cm
% Give a unique label
\caption{Density plot of the bipartite entanglement (a) $E_{\mathcal{N}\text{%
,}am}$, (b) $E_{\mathcal{N}\text{,}bm}$ and (c) $E_{\mathcal{N}\text{,}ab}$
versus the detuning $\Delta $ and the ratio of $G/g_{ma}$ ($g_{ma}$ is
fixed). (d) Density plot of the bipartite entanglement $E_{\mathcal{N}\text{,%
}am}$ versus $G/g_{ma}$ ($g_{ma}$ is fixed) and $\protect\kappa _{a}/\protect%
\kappa _{m}$ ($\protect\kappa _{m}$ is fixed). The parameters are $\protect%
\kappa _{a}=0.2\protect\kappa _{m}$, $g_{ma}=\protect\omega _{b}$, and the
other parameters we chose are the same as those in Fig.2.}
\end{figure}

In Fig. 4(a), (b) and (c), we exhibit three bipartite entanglements $E_{%
\mathcal{N}\text{,}am}$, $E_{\mathcal{N}\text{,}bm}$ and $E_{\mathcal{N}%
\text{,}ab}$ vary with the detunings $\Delta $ and the ratio of $G/g_{ma}$.
As $G$ increases, the bipartite entanglements $E_{\mathcal{N}\text{,}am}$, $%
E_{\mathcal{N}\text{,}bm}$ and $E_{\mathcal{N}\text{,}ab}$ all increases. It
can be clearly found that with the enhancement of magnomechanical
interaction, the indirect photon-phonon coupling caused by magnons will also
generate entanglement, and the entanglement caused by indirect coupling is
larger than that caused by direct coupling, it is as mentioned in \cite{k7}.
In addition, it shows that around $\Delta \approx -\omega _{b}$, the
detuning is resonant with the mechanical sideband, the maximum entanglement $%
E_{\mathcal{N}\text{,}am}$ can be obtained. From Fig. 4(d), it shows the
entanglement $E_{\mathcal{N}\text{,}am}$ is enhanced as the system
approaches the gain-loss balance.

Fig. 5 displays the effect of $\mathcal{PT}$-symmetry on the Gaussian
quantum steering. We find that the one-way quantum steering is obtained by
introducing the $\mathcal{PT}$-symmetry. For the conventional CMM system,
there is no quantum steering. When the system is in unbroken $\mathcal{PT}$%
-symmetry regime and the driving field is adjusted to make $G$ meet the
required condition, there exist entangled states which are $m\rightarrow b$
and $a\rightarrow b$ one-way steerable. It is not shown in Fig. 5 that $%
S_{b\rightarrow m}$ and $S_{b\rightarrow a}$ are both zero under different $G
$. The one-way quantum steering indicates that Bob can convince Alice that
the shared state is entangled, while the converse is not true. Its
application is that it provides one-side device independent quantum key
distribution (QKD), where the measurement apparatus of one party only is
untrusted, and it is has been experimentally observed \cite{k50,k51}.

\begin{figure}[tbp]
\centering% Use the relevant command for your figure-insertion program
\resizebox{0.45\textwidth}{!}{  \includegraphics{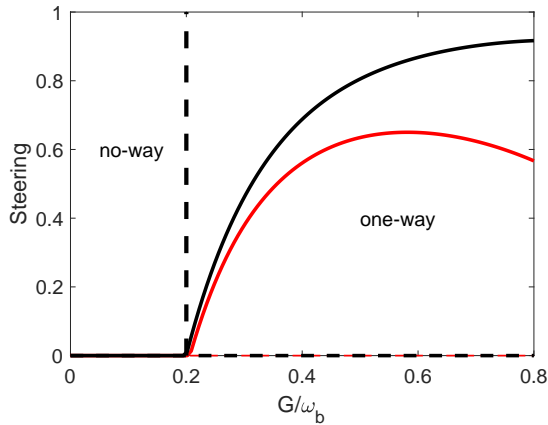}
} % If not, use
%\vspace{5cm}       % Give the correct figure height in cm
% Give a unique label
\caption{The quantum steerings $S_{m\rightarrow b}$ and $S_{a\rightarrow b}$
as functions of the effective coupling rates $G$. The black solid line
(dotted line) denotes $S_{m\rightarrow b}$ in the $\mathcal{PT}$-symmetric
CMM system (conventional CMM system), and the red solid line (dotted line)
denotes $S_{a\rightarrow b}$ in the $\mathcal{PT}$-symmetric CMM system
(conventional CMM system). We set $g_{ma}=\protect\omega _{b}$ and $\protect%
\kappa _{a}=$ $\pm 0.2\protect\kappa _{m}$, the other parameters we selected
are the same as those in Fig.2}
\end{figure}

It is important to discuss the influence of thermal noise on the
entanglement for the quantum devices. Fig. 6 shows the robustness of the
entanglement and steering against the temperature, we have plotted the
entanglement and steering as the functions of the temperature $T$. In Fig.
6(a), both the entanglement and steering are discussed in unbroken $\mathcal{%
PT}$-symmetry regime, it can be seen that the robustness of $E_{\mathcal{N}%
\text{,}am}$ is better than that of $E_{\mathcal{N}\text{,}bm}$ and $E_{%
\mathcal{N}\text{,}ab}$, and it survives up to 180mK. The robustness of
steering $S_{m\rightarrow b}$ and $S_{m\rightarrow b}$ are similar to that
of $E_{\mathcal{N}\text{,}bm}$, and they all survive up to about 40mK. Then
we show $\mathcal{PT}$-symmetry enhances the robustness of entanglement in
Fig. 6(b), the maximum temperature of entanglement $E_{\mathcal{N}\text{,}am}
$ is increased from about 147mK to 190mK.

Before ending this section, the experimental implementation is discussed. In
this $\mathcal{PT}$-symmetric cavity magnomechanical system, the
magnon-drive detuning can be adjusted not only by changing the frequency of
the microwave driving field, but also by the frequency of the uniform magnon
mode, which is modulated by an adjustable bias magnetic field $H$ in the
range of 0 to 1T \cite{k8}. And the magnon-photon coupling can be well tuned
by adjusting the bias magnetic field \cite{k46}. In addition, because of the
material characteristics of microwave cavity, we assume that the active
cavity mode can be construct by doping the active metamaterials with
inherent enhancing into the cavtiy. And the assumption is based on two
existing work: (1) the $\mathcal{PT}$-symmetric whispering-gallery
microcavities are achieved in the experiment \cite{k18}. (2) a system can
realize the acoustic gain by nonlinear active acoustic metamaterials \cite%
{k47}. For the quantum system we considered, the bipartite entanglement can
be obtained by measuring the cavity field quadratures. The cavity field
quadratures can be measured directly by homodyning the cavity output, and
the magnon state can be measured indirectly by homodyning the cavity output
of a introduced probe field. Moreover, we can used an additional optical
cavity to couple the YIG sphere, so that the mechanical quadratures can be
read out \cite{k7,k44}.

\begin{figure}[tbp]
\centering% Use the relevant command for your figure-insertion program
\resizebox{0.45\textwidth}{!}{  \includegraphics{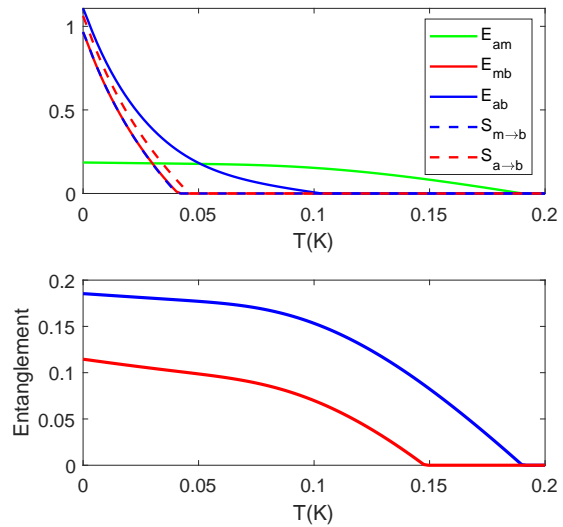}
} % If not, use
%\vspace{5cm}       % Give the correct figure height in cm
% Give a unique label
\caption{(a) The entanglements $E_{\mathcal{N}\text{,}am}$, $E_{\mathcal{N}%
\text{,}bm}$, $E_{\mathcal{N}\text{,}ab}$ and steering $S_{m\rightarrow b}$, 
$S_{a\rightarrow b}$ as functions of the temperature $T$, $\protect\kappa %
_{a}=$ $0.2\protect\kappa _{m}$. (b) The entanglements $E_{\mathcal{N}\text{,%
}am}$ as functions of the temperature $T$. The blue (red) solid line denotes
the case of $\mathcal{PT}$-symmetric CMM system $\protect\kappa _{a}=$ $0.2%
\protect\kappa _{m}$ (conventional CMM system $\protect\kappa _{a}=$ $-0.2%
\protect\kappa _{m}$). We set $g_{ma}=\protect\omega _{b}$, the other
parameters we selected are the same as those in Fig.2}
\end{figure}

\section{Conclusions}

In summary, we have investigated the enhancement of the bipartite
entanglement in a $\mathcal{PT}$-symmetric CMM system. By calculating linear
stability of the system, we find that the stability of the system decreases
with the system approaches the gain-loss balance. Compared with the cases of
broken $\mathcal{PT}$-symmetry and conventional CMM systems, the unbroken $%
\mathcal{PT}$-symmetric system is more stable in the range of parameters we
consider. Then we show that the bipartite entanglement and the robustness of
entanglement against environmental temperature are obviously enhanced by $%
\mathcal{PT}$-symmetry through comparison of the conventional CMM system. By
selecting appropriate driving field, one-way quantum steering between
magnon-phonon and photon-phonon modes can be observed by introducing $%
\mathcal{PT}$-symmetry. The experimental implementation is also discussed.
We believe that the proposed scheme provides a method for the entanglement
generation and the control of quantum steering in present cavity
optomechanics and it has potential applications in quantum optical devices
and quantum information networks.

\section*{ACKNOWLEDGEMENTS}

We thank Y. X Zeng for his fruitful discussion. This work was supported by
National Natural Science Foundation of China (NSFC): Grants Nos. 11574041
and 11375037.

\bibliography{08102020.bib}
% Produces the bibliography via BibTeX.

\end{document}